\documentclass[10pt,journal,letterpaper,compsoc]{IEEEtran}

\usepackage{amsmath}

\usepackage{epsfig,subfigure}
\usepackage[ruled,vlined]{algorithm2e}

\providecommand{\eat}[1]{}

\begin{document}
\title{{Mixing MACs: An Introduction to Hybrid Radio Wireless Virtualization}}

\author{\authorblockN{Gautam Bhanage\authorrefmark{2}} \\
\authorblockA{\authorrefmark{2}Tech Report: Bhanage.com / GDB2017-005} \\
gbhanage@gmail.com}

\maketitle

\begin{abstract}
This study presents the design of the hybrid wireless virtualization (\textbf{HWV}) controller based network architecture. Using a HWV controller, an  unified approach can be taken for provisioning and management of  virtualized heterogeneous radios, irrespective of their MAC and PHY layer mechanisms. It is shown that the airtime occupancy by transmissions from different slices or groups can be used as a single metric for tying these virtualized platforms. The HWV controller can account and dynamically re-provision slice quotas, which can be used for maximizing the network operator's revenue or aggregate system throughput performance. Results from simulations show that an HWV controller based infrastructure is able to improve the revenue generated from a single virtualized basestation and an AP by up to $40\%$ under tested conditions.
\end{abstract}


%
\IEEEpeerreviewmaketitle
 
\section{Introduction}
\PARstart{W}ireless virtualization can be defined as the approach by which the individual underlying physical network interfaces are abstracted by more than one virtual interfaces, which allow for better sharing of the physical interface and the medium used by the radio associated with that interface. The term virtualization itself was coined from the server systems area of virtualization and usually involves the application of three fundamental concepts for sharing the resource: (1) abstraction, (2) programmability, and (3) isolation. \emph{Abstraction} ensures that the same interface is available to all the entities using the systems. \emph{Programmability} ensures that each of the virtual entities (interfaces in our case) are able to affect properties of the underlying physical interface without conflicting with the requirements of other virtual entities. Finally, \emph{isolation} is a property which ensures that the load on one virtual entity does not affect the other, and essentially, each of the interfaces are able to work oblivious of each other. In our wireless case, virtualization refers to running multiple virtual networks which are leased to mobile virtual network operators (MVNOs) by the mobile network operator (MNO) which owns the underlying physical networks.

\begin{figure}[t]
\begin{center}
\epsfig{figure=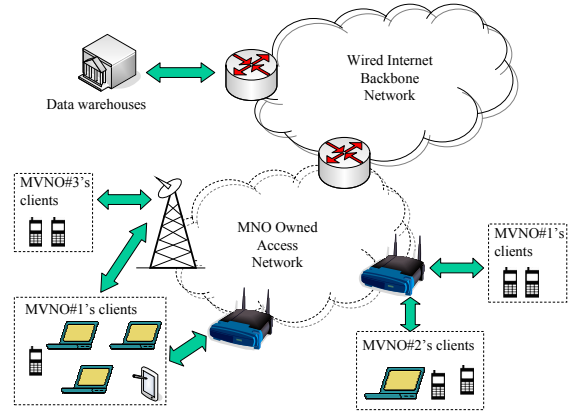,width=2.9in}
\caption{An example of a virtualized network architecture. The figure shows two slices belonging to MVNO1 and MVNO2 which are hosted by the network operator. These slices are hosted across different virtualized wireless substrates.} \label{fig:virtualized_net_arch_2}
\end{center}
\end{figure}

Recent virtualization wireless efforts have shown how individual wireless network interfaces like those in a 4G cellular basestation~\cite{Bhanage10Virtual,bhanage2012virtualization} and a cellular WiFi hotspot~\cite{Bhanage11:SplitAP} can be virtualized. However, in this study we wish to look at a scenario beyond individual virtualized wireless components, where a central framework could be employed by a network operator for maximizing its revenue from the MVNOs. Consider the virtualized network architecture shown in Figure~\ref{fig:virtualized_net_arch_2}. As seen in the figure, a MNO will have access to different types of virtualized radio platforms, and a mechanism is needed for establishing control and provisioning across these different radios. Our HWV controller design will be able to leverage such configurable virtualized radio platforms for revenue maximization of the MNO.

The contributions\footnote{The ideas presented here are based on our issued patent~\cite{bhanage2013hybrid} at one of my previous employers. This work has no affiliation or relation to my current or other past employers. Opinions and thoughts discussed here are my own and no one else. Please use this document to stoke your thinking and none of these should be construed as advice.} made by this study can be summarized as follows:
\begin{itemize}
\item{Metric: }We discuss and show how a single metric could be used for addressing the issue of resource accounting and allocation across a diverse set of radio interfaces.
\item{Framework: } We present a mathematical optimization framework based on the above metric that will allow for enabling revenue ($\$$/sec) or capacity maximization (bits/sec) of the physical network by dynamically reprovisioning individual virtualized wireless networks or interfaces.
\item{Prototype design:} Guidelines are laid out for our work-in-progress HWV controller based network architecture design.
\item{Simulations: } Finally, using simulations, we show some preliminary results that can be obtained by deploying our HWV controller with virtualized wireless network systems.
\end{itemize}

Rest of this document is structured as follows. Section~\ref{sec:related} discusses related work. Section~\ref{sec:hwv_design} presents our hybrid wireless virtualization based network architecture and platform. Section~\ref{sec:simulations} presents results from our simulation of the controller with a virtualized AP and basestation. Finally, Section~\ref{sec:conc} presents conclusions and future directions.

\section{Related Work}
\label{sec:related}

Previous network virtualization~\cite{wen2013wireless, liang2015wireless, bhanage2011network} efforts can be broadly classified as those for wired networks and for wireless networks. In terms of wired networks, virtualization is used for running testbeds~\cite{peterson06,VINI}, and also for dynamically re-configuring and maintaining routers~\cite{vroom}.  A host of other studies have discussed how topologies can be managed across virtualized wired networks~\cite{rethinking09rexford,fan06dynamictopology}, and some also for wireless networks~\cite{Bhan1109:Virtual}.

Wireless device virtualization was first  proposed for short range WiFi radios for virtual access points(VAPs)~\cite{aboba}. These VAPs would run as abstractions on a physical AP.  One of the first virtual basestation designs is the VANU MultiRAN virtual basestation design~\cite{vanu}. It runs the entire virtualized basestation transceiver (BTS) stack from software. A similar approach is considered for the Open Basestation project~\cite{openbts}which relies on implementing a 2G BTS in software, though it does not support virtualization. Another study~\cite{miguel08_3gpp} discusses approaches in which the network architecture may be emulated using virtual machines, but does not deal with the emulation of the radio itself. An approach that modified a proprietary MAC scheduler for virtualizing a BTS is discussed in another previous work~\cite{Kokku10NVS}.  Another of our previous works has focused on mapping virtual wireless networks to physical wireless networks~\cite{bhanage2011virtual}. However, None of these previous studies discuss a unified approach to managing virtualized wireless networks, which is the focus of this study.

Our work shows how a network architecture with heterogeneous virtualized radios will work with the help of some of our previous virtualization frameworks. An approach based on virtualizing and customizing a carried grade basestation is discussed in the virtual basestation~\cite{Bhanage10Virtual} (vBTS) study, and this will be controlled by a hybrid wireless virtualization controller, which is discussed later. In terms of access points, a modified version of our previous work on SplitAP~\cite{Bhanage11:SplitAP} which virtualizes and provisions airtime across VAPs will be leveraged. The reason for choosing our framework for designing a hybrid virtualized wireless network are two fold. (1) Our framework does not involve any proprietary components, and can be employed on commodity hardware, and (2) our framework is capable of using airtime as a means for accounting and control. The importance of using airtime will be discussed in further sections.

\section{HWV Design}
\label{sec:hwv_design}

We begin our discussion of the HWV design with a formalization of the design goals of the architecture, followed by metric selections, the HWV model, and finally an example deployment discussion.

\subsection{Architectural Goals}
The broad design goals of a HWV system can be further crystalized into the following requirements:
\begin{itemize}
\item The network architecture must support virtualization and resource allocation across multiple wireless endpoints, such as basestations, that are owned by the network operator. This support should be independent of the type of the wireless network hardware itself.  For example, they could be 2G basestations, 3G basestations, or WiFi hotspots.
\item The hybrid wireless virtualization (HWV) architecture should be able to dynamically re-provision and control the virtualized network such that the network operator benefits irrespective of time varying operating conditions.
\end{itemize}
The first goal has been independently addressed in prior studies~\cite{Bhanage10Virtual}\cite{Bhanage11:SplitAP}. In order to ensure that the second goals is met, we propose building a HWV controller that uses a single metric for measurements and policy enforcements across multiple virtualized radios owned by the network operator.

\subsection{Metric Selection}

For the purpose of link and user performance measurement, there are several widely used metrics like the throughput, goodput or even the delay performance experienced by the user. Due to their popularity and direct impact on the end user experience, these metrics are also used in resource provisioning on scheduled MAC mechanisms like 4G basestations. For example, in a WiMAX basestation, the system administrator can specify the number and types of flows in a service class, with each class having a minimum, maximum throughput and delay constraints associated with it. These help the basestation schedule links to better suit the application requirements. However, these metrics cannot be used for our HWV architecture. The main reasons for not being used in our framework are that these metrics do not translate into the use of the same amount of radio resources on any type of radio. For example, the same amount of throughput can be achieved by using different physical layer rates on two different links. However, the link with the slower physical layer rate uses more radio resources to achieve the same throughput as that by the faster radio link. In order to remove these, and other differences in resource usage induced by the MAC layer, such as channel access time, MAC and PHY overheads, MAC enhancements (like custom or standard aggregation mechanisms~\cite{bhanage2017amsdu}) and the nature of the MAC itself (scheduled or unscheduled), we propose using radio airtime as a single metric for control of the virtualized framework. Note that earlier studies~\cite{Kokku10NVS} have maintained that rates for a single radio can be used as alternatives. However, as will be seen through further discussion, fairness across radio technologies can be truly represented through airtime fairness.

\noindent\textbf{Universality of $t_j$}: The fraction of airtime $t_j$ used by any radio for wireless transmission directly translates into the resources on every wireless device. In a previous work, it has been shown that the reservation rate in a WiMAX scheduler can be used for specification of required time slots.  We will further show that airtime can be used as a single metric for resource accounting on a scheduled as well as unscheduled MAC. Consider that $\mu_j$ represents the number of resource blocks in a 4G basestations MAC scheduler that are allocated to slice $j$. In this case, it can always be experimentally verified that:
\begin{equation}
t_j = \frac{\mu_j}{\sum_j \mu_j} = \frac{rate_j}{rate^{phy}_j},
\end{equation}
This is the case because airtime utilization of a slice is a direct result of such MAC scheduling. In any radio, the airtime fraction is also equivalent to the aggregate throughput $rate_j$ achieved by slice $j$, as a fraction of the assigned average physical $rate^{phy}_j$ for the slice. Hence, irrespective of having a scheduled MAC (in a basestation) or an unscheduled CSMA MAC (in an AP), airtime can be used as a single number for accounting and controlling slice radio resource usage. 

With the advent of faster WiFi mechanisms like 802.11ac and 802.11ax, this metric should provide a way to account for MU$-$MIMO and OFDMA mechanisms. If the underlying OFDMA radio is divided on subcarriers, then this can still be translated into airtime by normalizing the airtime utilized by the number of subcarriers alloted over a period of time. Similarly, for MU$-$MIMO radios, the airtime can be normalized by the number of simultaneous transmissions to ensure that we account for fairness between MU and non$-$MU transmissions. 

Based on this insight, we will now present the design of the HWV controller that will take into account resource usage across different virtualized network components, and will be responsible for dynamically reprovisioning these components for network operator revenue maximization or plain rate maximization.

\subsection{HWV Controller Model}
\noindent\textbf{Formulation:}  The HWV controller model is responsible for getting airtime usage, and rate feedback from each of the virtualized components such as basestations and AP. It is also responsible for reprovisioning them so that the network operator's profit is maximized. The virtual basestation~\cite{Bhanage10Virtual} architecture and the modified SplitAP~\cite{Bhanage11:SplitAP} framework enforce group airtime fairness across slices. 

Let the airtime quota requested per
slice across a group of basestations, or access points be specified
by the set $Q = \{Q_1, \ldots Q_n\}$. Let the airtime allocated per
slice $j$ at every basestation $k$ be given as $t^{k}_j$. Similar
quotas $\hat{Q} = \{\hat{Q_1}, \ldots \hat{Q_n}\}$ are specified for
the same set of slices over a set of access points. Let the airtime
allocated per slice $j$ at every access point $k$ be given as
$\hat{t^{k}_j}$. Similarly, let the requested airtime per slice per
basestation or access point be given as $r^{k}_j$, and
$\hat{r^{k}_j}$ respectively. The requested airtime is updated at
runtime for every iteration at which the optimization problem is
solved. Corresponding usage flags for basestation and access points
are given as: $u^{k}_j$, and $\hat{u^{k}_j}$.

We define the overall revenue function across the set of
basestations and access points owned by the operator as:
\begin{equation}
Rev(t^k_j) =   t^{k}_j \times \Gamma_j(r^{k}_j)
\end{equation}

\begin{equation}
\hat{Rev(\hat{t^k_j)}} =  \hat{t^{k}_j} \times
\hat{\Gamma_j}(\hat{r^{k}_j})
\end{equation}
Here, $\Gamma_j$ and $\hat{\Gamma_j}$ are the utility functions
provided by MVNO $j$ for matching their traffic demands on a
basestation and access point respectively. For now, we can define
the utility functions as a linear function of the allocated airtimes
given as:
\begin{equation}
\Gamma_j(t^{k}_j) =  C_j
\end{equation}
\begin{equation}
\hat{\Gamma_j}(\hat{t^{k}_j}) = \hat{C_j}
\end{equation}

These equations indicate a purely increasing utility with increasingly allocated capacity. Eventually, the values $C_j$ and $\hat{C_j}$ can be equated to the average physical rates available for the clients belonging to the slices $j$ on basestations and access points. Information of the average physical layer rates, and the overall traffic flowing to the clients is available at the controllers of both the virtual baseation framework and the SplitAP design and these can be polled regularly by the HWV controller. In our initial evaluation of the setup, we will set $C_j$ and  $\hat{C_j}$ equal to average of the slice physical layer rates to the clients. This eliminates the pricing component from the objective function, and the problem becomes a rate maximization problem. Eventually, we demonstrate the revenue maximization function of the controller by substituting linear objective functions of achieved rate for  $C_j$ and  $\hat{C_j}$. The problem being solved at the controller can finally be formulated as:

\begin{equation*}
\begin{aligned}
& \underset{}{\text{maximize}}
& & \sum_j \sum_k Rev(t^k_j) + \hat{Rev(t^k_j)} \\
& \text{subject to} & & \sum_j t^k_j \leq 1, \; j = 1, \ldots, m.\\
&                   & & \sum_k t^k_j \leq Q_j, \; k = 1, \ldots, n.\\
&                   & & \forall_{j,k}\texttt{ }t^k_j \geq \delta^k_j, \; \\
&                   & & \sum_j \hat{t^k_j} \leq 1, \; j = 1, \ldots, m.\\
&                   & & \sum_k \hat{t^k_j} \leq \hat{Q_j}, \; k = 1, \ldots, n.\\
&                   & & \forall_{j,k}\texttt{ }\hat{t^k_j} \geq \hat{\delta^k_j}, \; \\
\end{aligned}
\end{equation*}
In the above optimization formulation, the $\delta^k_j$ is used to represent the minimum airtime reservation at the basestation $k$ for the slice $j$. Since the objective function, and all of the constraints are convex, the formulation can be solved at the HWV controller using any standard convex optimization tool or heuristic.

\subsection{Prototype Design}
In this section we will discuss the work in progress for building the HWV controller and network architecture prototype. For this proof of concept architecture, we leverage the previously designed virtual wireless basestation (vBTS) prototype~\cite{Bhanage10Virtual}, and a modified SplitAP~\cite{Bhanage11:SplitAP} based virtualized AP prototype that allows us to control downlink group airtime quotas. Both of these prototypes allow the network operator dynamically controller the slice quotas. Both the vBTS framework and the SplitAP framework are connected to the HWV controller running on the network through an IP backhaul.

\begin{figure}
\begin{center}
\epsfig{figure=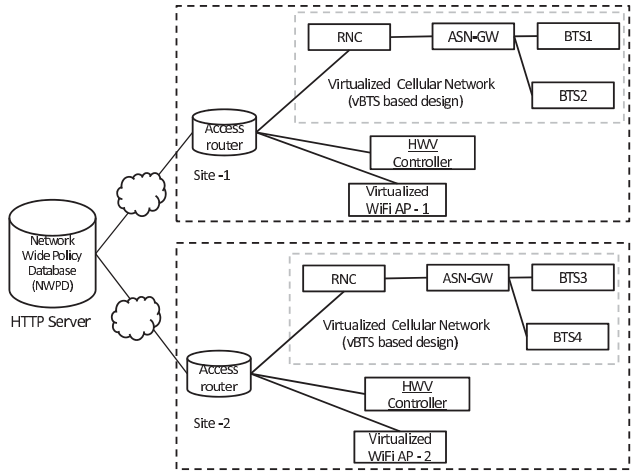, width=3.4in}
\caption{An example deployment of the HWV controller in a virtualized network architecture. Since the HWV controller is reachable by IP it can be placed wherever there is IP connectivity. However, for achieving fine grained control, it is better placed closer to the edge or near the cells that it controls. } \label{fig:nw_arch_proc}
\end{center}
\end{figure}

An example network layout is as shown in the Figure~\ref{fig:nw_arch_proc}. As seen we propose deploying independent HWV controllers in access networks, and each of these HWV controllers are responsible for controlling a limited set of vBTS frameworks and SplitAPs. For every operator, these independent HWV controllers in turn will be connected to a HTTP based network wide policy database (NWPD), that advertises the operators revenue generation capabilities through different access mechanisms like cellular links or WiFi hotspots. The NWPD will also be responsible for advertising utility functions for each of the slices. We have made a provision for advertising the utility functions because we envision that these will change based on time varying agreements between the network operator (MNO) and the virtual network operators (MVNOs). Each of the HWV controllers will be configured at deployment time with the URL of the NWPD, and will be responsible for independently fetching operator policies and utility functions. Note that the fetch from the site-local HWV controller can happen very infrequently, and is dependent on the time of the lease agreements between the MNO and the MVNO. On the other hand, the control loop between each of the HWV controllers and their connected substrates will be on a much finer scale.

%

\section{Simulations for HWV Evaluation}
\label{sec:simulations}

We present some initial results from our virtualization setup.

\begin{table}
\begin{center}
\caption{Simulation parameters.}
  \begin{tabular}{| l | l | }
    \hline
    Parameter & Value \\ \hline \hline
    Simulation runs & 1000 \\ hline
    APs & 1  \\ \hline
    BTSs & 1 \\ \hline
    AP peak throughput & 36Mbps \\ \hline
    BTS peak throughput & 20Mbps \\ \hline
    SLC1 airitme bid & 1.4 \\ \hline
    SLC2 airitme bid & 0.6 \\ \hline
    Slices per AP & 2  \\ \hline
    Slices per BTS & 2  \\
    \hline
  \end{tabular}
   \label{tab:simulation_params}
\end{center}
\end{table}

\subsection{Setup}
The setup consists of our centralized HWV controller that is connected to a virtualized access point~\cite{Bhanage11:SplitAP}, and virtualized WiMAX basestation~\cite{Bhanage10Virtual}. The parameters for the simulation are as described in the Table~\ref{tab:simulation_params}\footnote[1]{Numbers for throughput values of the BTS and AP are based on measurements on commercial hardware.}. In our preliminary evaluation we determine the performance of the system with the loads on each of the slices varying randomly. The change in load is based purely on the requirements for different airtimes to support the same amount of traffic, which is either due to the change in the link conditions to the access point or the basestation.

\subsection{Unconstrained Setup with Varying Slice Rates}
In our first experiment, we consider that the network administrator has placed no constraints on the quota allocations at either substrate (AP or the BTS). In this case we plot the weights or airtime allocation quotas computed by our architecture and the corresponding revenue generated. Note that in this experiment, the HWV controller is configured to purely solve a throughput maximization problem.

\begin{figure}[t]
\begin{center}
\epsfig{figure=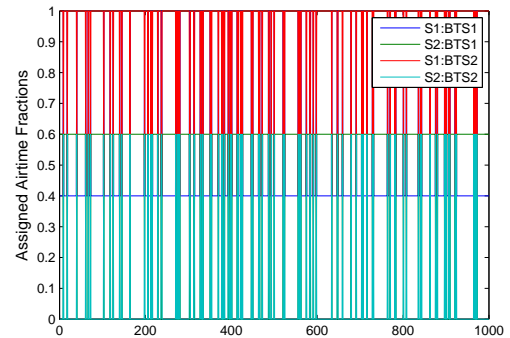,width=2.9in}
\caption{Weights (airtime allocations for slices) calculated by the HWV controller based on varying load conditions.} \label{fig:hwv_rate_wts}
\end{center}
\end{figure}

\begin{figure}[t]
\begin{center}
\epsfig{figure=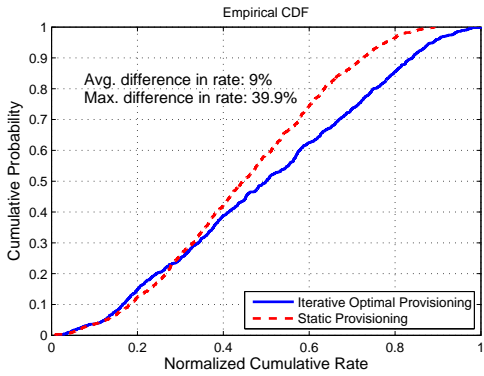,width=2.9in}
\caption{Revenue generated as a function of time by with the HWV controller as opposed to a static allocation.} \label{fig:hwv_rate_cdf}
\end{center}
\end{figure}

Figure~\ref{fig:hwv_rate_wts} shows the airtime allocations calculated by the HWV controller for the slices at the AP and the BTS based on the varying loads. The results show that based on the operating conditions, the HWV controller adapts the quota allocations at the substrates. These results are deliberately plotted as a function of time to show that per-cycle reconfiguration initiated by the HWV controller is capable of handling even random changes in loads. We observe that even though the quota allocations change significantly for each of the substrates, the patterns are not completely random since they are limited by the net airtime constraints for each of the slices, and the limitations of total airtime on each of the radios.

The CDF of the normalized rate generated from the corresponding experiment is as plotted in Figure~\ref{fig:hwv_rate_cdf}.  As seen in the results, we observe that on an average the revenue performance improves by at least $9.3\%$. In the best case, we observe that the revenue performance improve by up to $39.7\%$.

\subsection{Constrained Setup with Varying Slice Rates}

\begin{figure}[t]
\begin{center}
\epsfig{figure=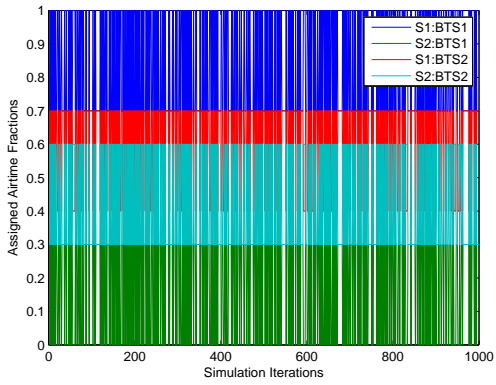,width=2.9in}
\caption{Weights (airtime allocations for slices) calculated by the HWV controller based on varying load conditions under additional airtime quota constraints.} \label{fig:hwv_rand_resv_w2}
\end{center}
\end{figure}

\begin{figure}[t]
\begin{center}
\epsfig{figure=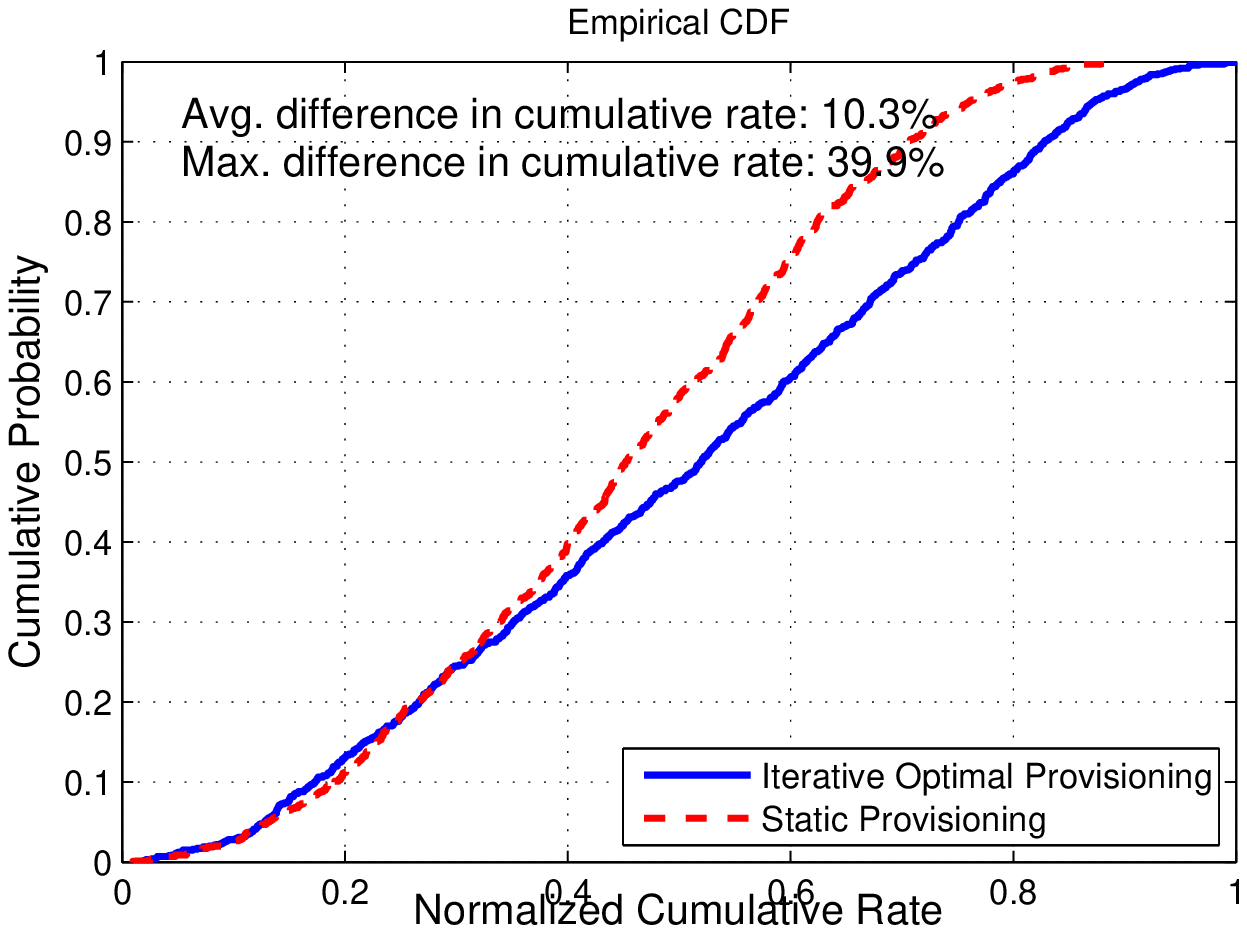,width=2.9in}
\caption{Revenue generated as a function of time by with the HWV controller as opposed to a static allocation under additional airtime quota constraints.} \label{fig:hwv_rand_resv_cdf}
\end{center}
\end{figure}

In the previous case, we measured performance of the HWV framework in the absence of any restrictions on the weights allocated to the individual substrates or the slices. Hence, in every case the controller, calculated a solution that maximizes aggregate rate across substrates without any external constraints and the results are influenced only by the utility functions for each of the substrates. However, our HWV controller allows us to impose constraints on the quotas allocated at each of the substrates, the AP and the basestation, for each of the slices. In this experiment, we impose a constraint of receiving a minimum allocation of 0.7 for the first slice on the AP.

Airtime allocations from the experiment are as shown in the Figure~\ref{fig:hwv_rand_resv_w2}. We observe that the allocations change based on the load conditions. However, we observe that the oscillations are not as much as those seen in Figure~\ref{fig:hwv_rate_wts} because of the constraints imposed on the desired airtime quotas. The CDF of the rate allocations generated from the experiment are as shown in the Figure~\ref{fig:hwv_rand_resv_cdf}.  This plot shows the CDF of the revenue generated with our HWV controller as opposed to using a static allocation. Results show that even in this case, despite of the airtime quota constraints the average revenue improves by $10\%$, and the best case revenue improves by up to $40\%$.

\subsection{Revenue Performance}

\begin{figure}[t]
\begin{center}
\epsfig{figure=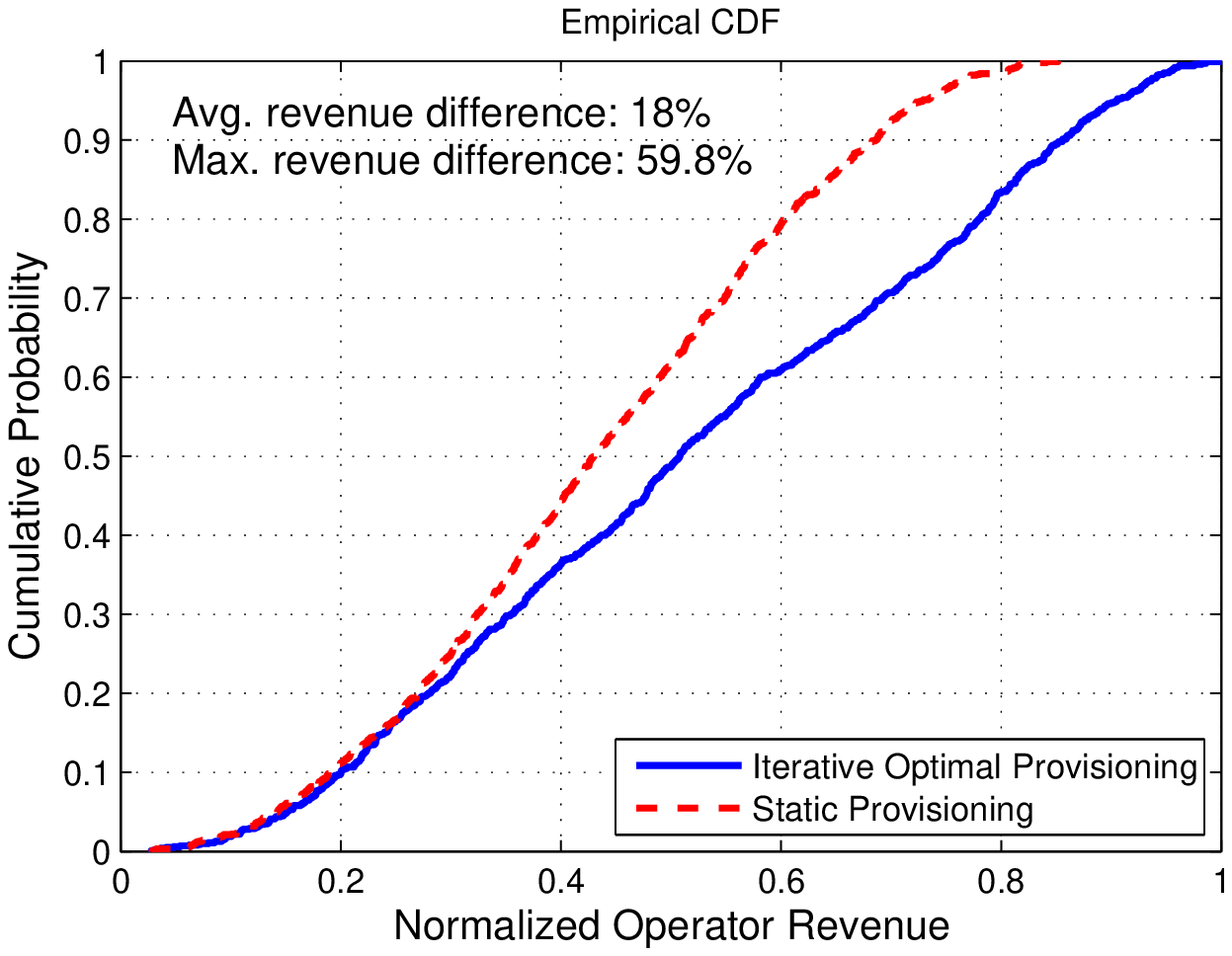,width=2.9in}
\caption{Revenue generated as a function of time by with the HWV controller as opposed to a static allocation under additional airtime quota constraints.} \label{fig:hwv_rate_price_cdf}
\end{center}
\end{figure}

In all of the previous cases, we have shown how our architecture can solve an aggregate rate maximization problem for the network operator when pricing is not involved. In this study we show how the operator could include pricing information in the HWV architecture and use it to solve a revenue maximization problem. In this experiment, we assume that the revenue generated out of the 4G WiMAX BTS is twice as that obtained from the WiFi AP. Accordingly, this information is used to condition the objective function in the controller. Rest of the experiment is as before, and results are measured for iterations with randomly generated average slice rates.

Results from this experiment are as shown in the Figure~\ref{fig:hwv_rate_price_cdf}. The results show a CDF of the revenue generated through a static allocation strategy and that obtained through a dynamic re-allocation approach using our HWV controller. The results show that adding our HWV controllers re-allocation strategy for changing average rate conditions, for differently priced substrates allows the operator to do significantly better than a static slice allocation. In this case, we observe that the average improvement in revenue is $10\%$, with a best case improvement of up to $40\%$.

\eat{
\subsection{Varying Slice Loads}

\begin{figure}[t]
\begin{center}
\epsfig{figure=hwv_load_w,width=2.9in}
\caption{hwv - Weights} \label{fig:hwv_resv_w}
\end{center}
\end{figure}

\begin{figure}[t]
\begin{center}
\epsfig{figure=hwv_load_rev,width=2.9in}
\caption{hwv - Revenue} \label{fig:hwv_rev_}
\end{center}
\end{figure}
} 

\section{Conclusions and Future directions}
\label{sec:conc}

This study presents the design and the initial evaluation of the hybrid wireless virtualization (HWV) controller mechanism. We show that the HWV controller operates across heterogenous virtualized radio technologies and maximizes network operator revenue based on dynamic airtime quota re-allocations.  We show that the objective for revenue maximization from multiple virtualized radio interfaces can be formulated as a convex optimization problem, under the conditions that the utility functions from each of the virtualized radios are convex. Results from preliminary evaluations show that in certain cases our setup can be improve revenue by up to $39\%$ over static allocation schemes. The average performance improvement in an unconstrained case, with no minimum quota limitations for the slices  is seen to be approximately $40\%$. Though the absolute numbers in specific cases will possibly vary based on load conditions, physical layer rates, and actual radio efficiencies, we see that our HWV controller is able to successfully improve the aggregate network rate or the revenue for the network operator by dynamically changing airtime allocations for slices at run time.

\bibliographystyle{IEEEtran}

\end{document}